\documentclass[aps,preprint]{revtex4}%
\usepackage{amsfonts}
\usepackage{amsmath}
\usepackage{amssymb}
\usepackage{graphicx}%
\begin{document}
\title{Charge Tempered Cosmological Model}
\author{Mario Goto (mgoto@uel.br)}
\affiliation{Universidade Estadual de Londrina}

\begin{abstract}
The main purpose of this work is to obtain the metric of a Charge Tempered
Cosmological Model, a slightly modified Standard Cosmological Model by a small
excess of charge density, distributed uniformly in accordance with the
Cosmological Principle, the global Coulomb interaction incorporated in this
metric. The particularity of this model is that the commoving observer
referential where the metric belongs is non inertial, which consequence is
that clocks at different position can not be synchronized. The new metric is
constrained to $k=0$, with dependence on a charge parameter, and related to a
modified Friedmann equation, but it is constrained to a positive deceleration
parameter and hyperbolic solution. Nevertheless, there are corrections to do,
valid just for a long range distances. For example, the red shift has, now,
dependences on the gravitational potential together the recessional motion. In
any way, this model accepts as well the cosmological constant and its physical
counterpart, the dark energy.

\end{abstract}
\maketitle

\section{Introduction}

It is a common sense that, among the two long range interaction we know,
gravitational and electromagnetic, only the first, being universal and
cumulative, can be responsible for the great scale cosmic dynamic. So, the
Standard Friedmann and Robertson-Walker Cosmological Model $^{[1-3]}$ uses as
a recipe an universe with an uniform distribution of matter and energy, which
the dynamic given solely by the gravitational interaction obeying Einstein
equations, except at the primordial Big Bang inflation, where all the
interaction is believed to be unified and obeying the Quantum Mechanics laws
$^{[4]}$. As a result, we have an evolutionary, actually expanding universe,
which must be decelerated due to the gravitational attractive interaction.

However, systematic observations of the recession velocities of type Ia
supernova indicated a positive acceleration (or negative deceleration
parameter) of the universe expansion$^{[5-9]}$

. To take account to these observations, the cosmological constant was
reintroduced in the Einstein equations, which gives us then the positively
accelerated solutions $^{[10-12]}$. Physically, addition of the cosmological
constant is equivalent to add an uniform and constant energy distribution with
negative pressure responsible by the positive acceleration. It is known as
dark energy, a counterpart of the dark matter used to solve the problem of
stelar rotation in galaxies. The nature of the dark energy, as well as the
dark matter, is not well understood, while the dark energy must be of the
order of 70 per cent of the whole energy content of the universe, the dark
matter 25 per cent and the normal matter just 5 per cent.

A natural way to introduce a repulsive acceleration is the Coulomb interaction
between charged particles, and there are many attempts to deal with, but
actual physics theories are strongly based upon symmetries and conservation
laws, and one of the most stringent is the charge conservation. It is
generally accepted that the universe as a whole have to be neutral because
there is no mechanism for charge production in any actual theories like the
Standard Model and other unified theories $^{[13-15]}$. The electromagnetic
interaction $^{[16]}$, due to the presence of opposite charges, positive and
negative, in perfect balanced amount, as it is believed, shouldn't be relevant
at the great cosmological scale. There are many works to show how much charge
asymmetry is admissible to be in agreement with present data $^{[17-26]}$.

Actually, we are not going to be stressed by such considerations because it is
reasonable to think that they depend on the geometrical environment, which is
considered as given by the Robertson-Walker metric
\begin{equation}
ds^{2}=-d\tau^{2}=-dt^{2}+R^{2}(t)\left[  \frac{dr^{2}}{1-kr^{2}}+r^{2}%
d\theta^{2}+r^{2}\sin^{2}\theta d\varphi^{2}\right]  \ , \label{RWalker}%
\end{equation}
the building base of the Standard Cosmological Model. We are using the metric
$g^{\mu\nu}$ compatible with the Minkowskian metric $\eta^{\alpha\beta}$ with
$diag(\eta^{\alpha\beta})=(-1,1,1,1)$.

The first thing we have to do is to conciliate the geometrical environment
with the physics content. If the physics content includes charge excess, the
geometrical environment should be changed because we are adding global non
gravitational interaction in such a way that the commoving referential frame
is not a free fall anymore.

Charged gravitational systems, in the context of General Relativity, has been
objects of recent studies, specially for isotropic and with rotational
symmetric systems as stars and black holes. In such studies, the
electromagnetic energy is incorporated to the total energy and momentum
tensor, and it contributes to the metric tensor to take account this
additional gravitational interaction source. However, Coulomb interaction is
not considered at all but for its global effects that cause instability of the
system until the great part of the charge excess had got away, as in the
charged black hole formation $^{[27-31]}$.

The region exterior to a spherically symmetric system with mass $M$ and charge
$Q$ is described by the Reissner-Nordstrom metric,
\begin{equation}
d\tau^{2}=\left[  1-\frac{2MG}{r}+\frac{Q^{2}G}{r^{2}}\right]  dt^{2}-\left[
1-\frac{2MG}{r}+\frac{Q^{2}G}{r^{2}}\right]  ^{-1}dr^{2}-r^{2}d\theta
^{2}-r^{2}\sin^{2}\theta d\varphi^{2}\ . \label{Reissner}%
\end{equation}
In the absence of charge, $Q=0$, and obviously it remits to the well known
Schwarzschild metric,
\begin{equation}
d\tau^{2}=\left[  1-\frac{2MG}{r}\right]  dt^{2}-\left[  1-\frac{2MG}%
{r}\right]  ^{-1}dr^{2}-r^{2}d\theta^{2}-r^{2}\sin^{2}\theta d\varphi^{2}\ .
\label{Schwarz}%
\end{equation}
A more generalized, taking into account a rotational symmetry, is the
Kerr-Newman metric, for systems with angular momentum.

To make clear the role of the charge content of the Reissner-Nordstrom metric
(\ref{Reissner}), let us go to write the equation of motion of a particle of
mass $m$ and charge $q$ placed in a region of influence of this metric at
radial distance $r$,%

\begin{equation}
\frac{d^{2}x^{i}}{d\tau^{2}}+\Gamma_{\mu\nu}^{i}\frac{dx^{\mu}}{d\tau}%
\frac{dx^{\nu}}{d\tau}=f_{ext}^{i}\ ,
\end{equation}
where $f_{ext}^{i}$ is an external non gravitational force, actually the
electrostatic force due to the charge $Q$. In Newtonian approximation, we have%

\begin{equation}
m\frac{d^{2}x^{i}}{dt^{2}}+m\Gamma_{00}^{i}=F_{ext}^{i}\
\end{equation}
for
\begin{equation}
\Gamma_{00}^{i}=-\frac{1}{2}\eta^{ij}\frac{\partial g_{00}}{\partial x^{j}%
}=-\frac{1}{2}\frac{\partial g_{00}}{\partial x^{i}}\ ,
\end{equation}
where
\begin{equation}
g_{00}=-\left[  1-\frac{2MG}{r}+\frac{Q^{2}G}{r^{2}}\right]
\end{equation}
and $\eta^{ij}$ are the spatial components of the Minkowskian metric. As the
system is isotropic, the equation of motion reduces to its radial component
\begin{equation}
m\frac{d^{2}r}{dt^{2}}=\frac{m}{2}\frac{\partial g_{00}}{\partial r}%
+F_{ext}^{r}\ .
\end{equation}
From (\ref{Reissner}),
\begin{equation}
\frac{1}{2}\frac{\partial g_{00}}{\partial r}=-\frac{MG}{r^{2}}+\frac{Q^{2}%
G}{r^{3}},
\end{equation}
and the equation of motion can be written as
\begin{equation}
m\frac{d^{2}r}{dt^{2}}=-m\left[  M-Q\phi(r)\right]  \frac{G}{r^{2}}%
+F_{ext}^{r}\ .
\end{equation}
It is clear the role of the charge component of the metric tensor as an
additional gravitational source due to the electrostatic self energy
$Q\phi(r)$, where
\begin{equation}
\phi(r)=-\frac{Q}{r}%
\end{equation}
is the electrostatic potential, and the Coulomb interaction force between
charges $q$ and $Q$ has to be put by hand. It is so because the charge term in
the metric (\ref{Reissner}) is just related to the additional electromagnetic
energy and momentum tensor put together with the matter source term of the
Einstein equations.

\section{\smallskip Cosmological Model}

Let us go to consider the Standard Cosmological Model slightly modified by an
uniform distribution of a small excess of electric charge, no matter where it
is from. In cosmological scale, there is no chance the charges scape to get
away. Instead, due to the electrostatic repulsion, any excess of charge will
be distributed uniformly in the whole universe and the Coulomb interaction
will act equally at all portion of the universe and certainly it is too much
stronger than the gravitational force due to its electromagnetic energy. At
this cosmological scale, matter and charge are constrained to move together
due to a combined gravitational and electrostatic forces, and therefore, as
the universe evolution is given by a scale factor, it is convenient to
incorporate this global electrostatic interaction in the structure of the
metric. It is the basic idea to build what one are going to name as the Charge
Tempered Cosmological Model.

Of course, the charge distribution depends on the nature of a possible charge
asymmetry, which can result in a charge distribution proportional to or
independent of the matter distribution. For example, if this eventual excess
of charges, positive or negative, is due to a proton-electron charge
asymmetry, the charge distribution is likely to be proportional to the matter
distribution. In the other hand, if it is from an asymmetry between the number
of positive and negative charged particles like the matter-antimatter
asymmetry, with a final excess of electrons in relation to protons or
vice-versa, due to the electrostatic repulsion and the extreme mobility of the
particles such as electrons, any excess of charge, positive or negative, will
be distributed uniformly, surpassing any matter inhomogeneity and getting a
complete annulation of the electromagnetic field, a scenario we are seeking for.

In an uniform universe, with matter, charge and anything else distributed
uniformly, there is no vector field, and scalar fields such as the
electrostatic potential must be spatially uniform. In such sense, Maxwell
equations are insensitive to an uniform charge distribution. Actually, it is
possible to think that the sources of Maxwell equations are related to the
fluctuations of the positive or negative charge distribution around a sea of
an uniform charge distribution that should be not necessarily neutral. From
the electromagnetic point of view, the assumption that the total charge
distribution of the universe is null may be so arbitrary as to assume that it
is not so.

\subsection{Commoving Referential}

The Standard Cosmological Model is built on the basis of the Cosmological
Principle, which postulates the homogeneity and isotropy of the universe,
geometrically traduced by the Robertson-Walker metric (\ref{RWalker}), where
the coordinates are defined on a commoving referential frame. It is locally
inertial and the time coordinate can be defined to coincide with the proper
time,
\begin{equation}
d\tau^{2}=-g_{\mu\nu}dx^{\mu}dx^{\nu}=-g_{tt}dt^{2}=dt^{2}\ ,
\label{propertime}%
\end{equation}
which permits the synchronization of all clocks at rest in this commoving
referential, and any object at rest will satisfy the free fall equation of
motion,
\begin{equation}
\frac{d^{2}x^{i}}{d\tau^{2}}+\Gamma_{\mu\nu}^{i}\frac{dx^{\mu}}{d\tau}%
\frac{dx^{\nu}}{d\tau}=\Gamma_{tt}^{i}=0\ . \label{comoving}%
\end{equation}

So, it implies
\begin{equation}
\Gamma_{00}^{i}=\frac{1}{2}g^{\mu i}\left(  \frac{\partial g_{0\mu}}{\partial
x^{0}}+\frac{\partial g_{0\mu}}{\partial x^{0}}-\frac{\partial g_{00}%
}{\partial x^{\mu}}\right)  =g^{ij}\frac{\partial g_{j0}}{\partial x^{0}}\ =0
\end{equation}
and, since $g^{ij}$ are components of a non singular matrix,
\begin{equation}
\frac{\partial g_{j0}}{\partial x^{0}}\ =0\ ,
\end{equation}
which is identically satisfied by the Robertson-Walker metric.

The absence of electrostatic field does not mean absence of electrostatic
interaction between parts of the system. Actually, each part of the system
acts, with a repulsive electrostatic force, on all the rest of the system, and
while the total force on it is null, it implies that the system as a whole is
under a nonzero pressure favor to increase the whole volume. Our present
challenge is to incorporate the global Coulomb interaction into the metric,
and we have to be ready to pay a just price for it. In a general case, it is
not possible to make the geometrization of the electromagnetic interaction as
it is done in gravitational interaction because the Equivalence Principle is
valid just for gravitation. However, in an uniform distribution of the charge
density in the whole universe, electromagnetic and gravitational forces, in a
large cosmological scale, must work together because there is just only one
degree of freedom to keep the homogeneity and isotropy of the Universe, and
the commoving referential is not any more a free fall locally inertial system.
As it acts globally, the Coulomb interaction can be incorporated into the
metric as the responsible for the referential acceleration.

Let us consider the equation of motion of a electrically charged particle with
charge $q$ and proper mass $m_{0}$,%

\begin{equation}
m_{0}\frac{d^{2}x^{\lambda}}{d\tau^{2}}+m_{0}\Gamma_{\mu\nu}^{\lambda}%
\frac{dx^{\mu}}{d\tau}\frac{dx^{\nu}}{d\tau}=qF_{\ \mu}^{\lambda}\frac
{dx^{\mu}}{d\tau}\ ,
\end{equation}
which spatial component is%

\begin{equation}
m_{0}\frac{d^{2}x^{i}}{d\tau^{2}}+m_{0}\Gamma_{\mu\nu}^{i}\frac{dx^{\mu}%
}{d\tau}\frac{dx^{\nu}}{d\tau}=f_{ext}^{i}\ ,
\end{equation}
where $f_{ext}^{i}$ is the Coulomb force that, in the commoving referential,
is $f_{ext}^{r}=qF_{\ \mu}^{r}U^{\mu}=qF_{\ 0}^{r}U^{0}$.

For the relativistic invariants $p^{\mu}p_{\mu}=-m_{0}^{2}c^{4}$ and
$g_{\mu\nu}U^{\mu}U^{\nu}=-1$, where $p^{\mu}=m_{0}U^{\mu}$, we have, in the
commoving referential, $g_{00}U^{0}U^{0}=-1$, so that
\begin{equation}
U^{0}=\frac{1}{\sqrt{-g_{00}}}\text{ and }\ U_{0}=\sqrt{-g_{00}} \label{U0}%
\end{equation}
and therefore
\begin{equation}
p^{\mu}=m_{0}\left(  U^{0},U^{i}\ \right)  =\frac{m_{0}}{\sqrt{-g_{00}}%
}\left(  c,v^{i}\ \right)  =m\left(  c,\mathbf{v}\ \right)  =\left(  \frac
{E}{c},\mathbf{p}\ \right)  .
\end{equation}

The radial equation of motion of a rest particle turn to be
\begin{equation}
\Gamma_{\mu\nu}^{r}\frac{dx^{\mu}}{d\tau}\frac{dx^{\nu}}{d\tau}=\frac
{f_{ext}^{r}}{m_{0}}=-\frac{q}{m\sqrt{-g_{00}}}E^{r}g_{00}U^{0}=-\frac
{q}{m\sqrt{-g_{00}}}E^{r}g_{00}\frac{1}{\sqrt{-g_{00}}}=\frac{q}{m}E^{r}\ ,
\end{equation}
where we are using the relativistic mass $m=m_{0}/\sqrt{-g_{00}}$ and we are
considering the isotropy.

So, in a commoving referential, any particle at rest must satisfy the
condition
\begin{equation}
\Gamma_{00}^{\mu}U^{0}U^{0}-\frac{f_{ext}^{\mu}}{m_{0}}=0 \label{comoving1}%
\end{equation}
or, considering radial symmetry,
\begin{equation}
-\frac{\Gamma_{00}^{r}}{g_{00}}-\frac{f_{ext}^{r}}{m_{0}}=0\ \label{comoving2}%
\end{equation}
as well as
\begin{equation}
-\frac{\Gamma_{00}^{0}}{g_{00}}-\frac{f_{ext}^{0}}{m_{0}}=0\ .
\label{comoving3}%
\end{equation}

For electrostatic force, due to the asymmetry of the electromagnetic tensor
$F^{\mu\nu}$,
\begin{equation}
f_{ext}^{0}=qF_{\ 0}^{0}\frac{dx^{0}}{d\tau}=0\ \label{comoving4}%
\end{equation}
such that we will have $\Gamma_{00}^{0}=0$.

From the affine connection
\begin{equation}
\Gamma_{\lambda\mu}^{\sigma}=\frac{1}{2}g^{\nu\sigma}\left\{  \frac{\partial
g_{\mu\nu}}{\partial x^{\lambda}}+\frac{\partial g_{\lambda\nu}}{\partial
x^{\mu}}-\frac{\partial g_{\mu\lambda}}{\partial x^{\nu}}\right\}  \ ,
\label{conection}%
\end{equation}
considering the isotropy, we obtain, from equations (\ref{comoving3})
and(\ref{comoving4}),
\begin{equation}
\Gamma_{00}^{0}=g^{r0}\left\{  \frac{\partial g_{0r}}{\partial x^{0}}-\frac
{1}{2}\frac{\partial g_{00}}{\partial r}\right\}  +\frac{1}{2}g^{00}%
\frac{\partial g_{00}}{\partial x^{0}}=0\ . \label{conection1}%
\end{equation}
Because it is possible, always, to impose $g_{00}$ be time independent, it
results
\begin{equation}
\Gamma_{00}^{0}=g^{r0}\left\{  \frac{\partial g_{0r}}{\partial x^{0}}-\frac
{1}{2}\frac{\partial g_{00}}{\partial r}\right\}  =0\ . \label{conection2}%
\end{equation}

From the spatial equation (\ref{comoving2}), we have
\begin{equation}
\Gamma_{00}^{i}=g^{ri}\left(  \frac{\partial g_{0r}}{\partial x^{0}}-\frac
{1}{2}\frac{\partial g_{00}}{\partial r}\right)  \neq0\ . \label{conection3}%
\end{equation}

Equations (\ref{conection2}) and (\ref{conection3}) are compatible only if
$g^{r0}=0$. Also, in a symmetric space, non diagonal components can be
eliminated with a redefinition of the coordinates. After these considerations,
we can go to construct the metric of our charge tempered universe.

\subsection{Charge tempered metric}

Let us consider the general form of the metric of an uniform space,
\begin{equation}
ds^{2}=-d\tau^{2}=g_{00}(r)dt^{2}+R^{2}(t)\widetilde{g}_{rr}(r)dr^{2}%
+R^{2}(t)r^{2}\left[  d\theta^{2}+\sin^{2}\theta d\varphi^{2}\right]  \ ,
\label{metric}%
\end{equation}
with the usual construction
\begin{equation}
g_{rr}(r,t)=R^{2}(t)\widetilde{g}_{rr}(r)\ , \label{grr}%
\end{equation}
where $R(t)$ is the space scale factor that carries the global time dependence
of the radial coordinate,%
\begin{equation}
r(t)=R(t)r\ . \label{R(t)}%
\end{equation}
.

Now, substitution of the component
\begin{equation}
\Gamma_{00}^{r}=-\frac{1}{2}g^{rr}\frac{\partial g_{00}}{\partial r}\
\end{equation}
of the affine connection in equation (\ref{comoving2}) leads to the condition
\begin{equation}
\frac{1}{2}\frac{g^{rr}}{g_{00}}\frac{\partial g_{00}}{\partial r}%
=\frac{f_{ext}^{r}}{m_{0}}=\frac{F}{m}\ , \label{condiction}%
\end{equation}
where $F$ is the radial Coulomb force that acts on any region of the universe.

Let us consider a spherical region with radius $r$ and charge $Q$, where
\begin{equation}
Q=\frac{4\pi}{3}\rho_{Q}r^{3}%
\end{equation}
for an uniform charge density $\rho_{Q}$. Now, let us take a small volume
$\Delta v$ with charge and mass contents, $q=\rho_{Q}\Delta v$ and
$m=\rho\Delta v$, respectively, where $\rho$ is the uniform mass density and
the small volume $\Delta v$ is placed on the surface of the spherical region.
The Coulomb force acting on $\Delta v$ due to its charge $q$ due to the charge
$Q$ is
\begin{equation}
F=\frac{qQ}{r^{2}}=\frac{4\pi}{3}\rho_{Q}^{2}\ r\Delta v\ ,
\end{equation}
corresponding to the force per mass unit
\begin{equation}
\frac{F}{m}=\frac{F}{\rho\Delta v}=\frac{4\pi}{3}\frac{\rho_{Q}^{2}}{\rho
}\ r\ .
\end{equation}
Substituted in (\ref{condiction}), it results%

\begin{equation}
\frac{1}{2}\frac{g^{rr}}{g_{00}}\frac{\partial g_{00}}{\partial r}=\frac{4\pi
}{3}\frac{\rho_{Q}^{2}}{\rho}\ r\ . \label{condiction1}%
\end{equation}

For a diagonal metric tensor, the orthogonality condition $g^{\mu\nu}%
g_{\nu\sigma}=g_{\sigma}^{\mu}=\delta_{\mu\sigma}$ implies $g^{\mu\nu}=\left(
g_{\mu\nu}\right)  ^{-1}$and, therefore $g^{rr}(r,t)=R^{-2}(t)\widetilde
{g}^{rr}(r)$, so equation (\ref{condiction1}) turn to be
\begin{equation}
\frac{1}{R^{2}}\frac{\widetilde{g}^{rr}}{g_{00}}\frac{\partial g_{00}%
}{\partial r}=\frac{8\pi}{3}\frac{\rho_{Q}^{2}}{\rho}\ r\ .
\label{condiction2}%
\end{equation}
After separation of temporal and spatial dependences, we have
\begin{equation}
\frac{1}{r}\frac{\widetilde{g}^{rr}}{g_{00}}\frac{\partial g_{00}}{\partial
r}=\frac{8\pi}{3}\frac{\rho_{Q}^{2}R^{3}}{\rho}\ =\lambda_{Q}\ ,
\end{equation}
where $\lambda_{Q}$ is the constant of separation of variables. The spatial
part, using $\widetilde{g}^{rr}(r)=\left(  \widetilde{g}_{rr}(r)\right)
^{-1}$, becomes
\begin{equation}
\frac{1}{g_{00}}\frac{\partial g_{00}}{\partial r}=\lambda_{Q}\ r\widetilde
{g}_{rr}\ , \label{g00}%
\end{equation}
and the temporal part defines the charge parameter
\begin{equation}
\lambda_{Q}=\frac{8\pi}{3}\frac{\rho_{Q}^{2}}{\rho}R^{3}\ . \label{LQ}%
\end{equation}

The metric functions $g_{00}(r)$ and $g_{rr}(r,t)$ must be obtained solving
the Einstein equations
\begin{equation}
R_{\mu\nu}=-8\pi GS_{\mu\nu} \label{Einstein}%
\end{equation}
together the constraint equation (\ref{g00}).

In principle, we must to add, explicitly, the electromagnetic energy
contribution to the source term of the Einstein equations. But the
electromagnetic energy must be uniformly distributed in such a way that it can
be incorporated to the matter density $\rho$, so the energy momentum tensor
will be the same perfect fluid of the Standard Model,%

\begin{equation}
T_{\mu\nu}=pg_{\mu\nu}+(p+\rho)U_{\mu}U_{\nu}\ . \label{Tmini}%
\end{equation}

The source term%

\begin{equation}
S_{\mu\nu}=T_{\mu\nu}-\frac{1}{2}g_{\mu\nu}T_{\ \lambda}^{\lambda}=\frac{1}%
{2}g_{\mu\nu}(\rho-p)+(p+\rho)U_{\mu}U_{\nu} \label{fonte}%
\end{equation}
has the time-time
\begin{equation}
S_{00}=-\frac{1}{2}(3p+\rho)g_{00} \label{S00}%
\end{equation}
and the three diagonal space-space
\begin{equation}
S_{ij}=\frac{1}{2}g_{ij}(\rho-p) \label{Sij}%
\end{equation}
as the non zero components.

Spatial homogeneity and isotropy impose the metric in the form (\ref{metric}),
its unknown components $g_{00}(r)$ and $g_{rr}(r,t)$ being connected by
(\ref{g00}), the energy and matter distribution given by (\ref{Tmini}) and the
constraint (\ref{LQ}) for charged matter distribution, all of then related by
Einstein equations (\ref{Einstein}).

The left side term of the Einstein equations (\ref{Einstein}) is the second
order Ricci tensor
\begin{equation}
R_{\mu\nu}=R_{\ \mu\lambda\nu}^{\lambda}\ , \label{Ricc}%
\end{equation}
contraction of the Riemann curvature tensor
\begin{equation}
R_{\ \mu\nu\kappa}^{\lambda}=\frac{\partial\Gamma_{\mu\nu}^{\lambda}}{\partial
x^{\kappa}}-\frac{\partial\Gamma_{\mu\kappa}^{\lambda}}{\partial x^{\nu}%
}+\Gamma_{\mu\nu}^{\eta}\Gamma_{\kappa\eta}^{\lambda}-\Gamma_{\mu\kappa}%
^{\eta}\Gamma_{\nu\eta}^{\lambda}\ , \label{Riemann}%
\end{equation}
fully dependent of the space-time geometry and its metric via affine
connection (\ref{conection}),
\begin{equation}
R_{\mu\nu}=\frac{\partial\Gamma_{\mu\lambda}^{\lambda}}{\partial x^{\nu}%
}-\frac{\partial\Gamma_{\mu\nu}^{\lambda}}{\partial x^{\lambda}}+\Gamma
_{\mu\lambda}^{\eta}\Gamma_{\nu\eta}^{\lambda}-\Gamma_{\mu\nu}^{\eta}%
\Gamma_{\lambda\eta}^{\lambda}\ . \label{Ricci}%
\end{equation}

In particular, the non zero components of the Ricci tensor are
\begin{equation}
R_{00}=\frac{\partial\Gamma_{0\lambda}^{\lambda}}{\partial x^{0}}%
-\frac{\partial\Gamma_{00}^{\lambda}}{\partial x^{\lambda}}+\Gamma_{0\lambda
}^{\eta}\Gamma_{0\eta}^{\lambda}-\Gamma_{00}^{\eta}\Gamma_{\lambda\eta
}^{\lambda} \label{R00}%
\end{equation}
and
\begin{equation}
R_{ij}=\frac{\partial\Gamma_{\mu\lambda}^{\lambda}}{\partial x^{j}}%
-\frac{\partial\Gamma_{ij}^{\lambda}}{\partial x^{\lambda}}+\Gamma_{i\lambda
}^{\eta}\Gamma_{j\eta}^{\lambda}-\Gamma_{ij}^{\eta}\Gamma_{\lambda\eta
}^{\lambda}\ . \label{Rij}%
\end{equation}

Exhaustive and systematic calculations are necessary to obtain, first, all
components of the affine connection, and then these Ricci tensor components.
Components of the affine connection are given in the appendix, from which we
can obtain the Ricci tensor. The time-time component (\ref{R00}) become%

\begin{equation}
R_{00}=3\frac{\overset{\circ\circ}{R}}{R}+\frac{3}{2R^{2}}\lambda_{Q}%
\ g_{00}+\frac{1}{4R^{2}}\left(  \lambda_{Q}\ r-\frac{\partial\widetilde
{g}^{rr}}{\partial r}\right)  \lambda_{Q}r\widetilde{g}_{rr}g_{00}%
\ \label{R00a}%
\end{equation}
which, using equation (\ref{g00a}), can be rewritten as
\begin{equation}
R_{00}=3\frac{\overset{\circ\circ}{R}}{R}+\frac{3}{2R^{2}}\lambda_{Q}%
\ g_{00}+\frac{1}{4R^{2}}\frac{1}{\widetilde{g}_{rr}}\frac{\partial g_{00}%
}{\partial r}\frac{\partial}{\partial r}\left[  \ln\left(  g_{00}%
\times\widetilde{g}_{rr}\right)  \right]  \,, \label{R00b}%
\end{equation}
where we are using the auxiliary notation
\begin{equation}
\frac{\partial R}{\partial t}=\frac{dR}{dt}=\,\overset{\circ}{R}\
\end{equation}
with large upper dot to indicate time derivative to distinguish it from the
proper time derivative with normal upper dot
\begin{equation}
\frac{\partial R}{\partial\tau}=\frac{dR}{d\tau}=\,\overset{.}{R}.
\end{equation}

The space-space component of the Ricci tensor, equation (\ref{Rij}), can be
decomposed as%

\begin{equation}
R_{ij}=\frac{\partial\Gamma_{i0}^{0}}{\partial x^{j}}-\frac{\partial
\Gamma_{ij}^{0}}{\partial x^{0}}+\left(  \Gamma_{i0}^{0}\Gamma_{j0}^{0}%
+\Gamma_{ik}^{0}\Gamma_{j0}^{k}+\Gamma_{i0}^{k}\Gamma_{jk}^{0}\right)
-\left(  \Gamma_{ij}^{0}\Gamma_{k0}^{k}+\Gamma_{ij}^{k}\Gamma_{0k}^{0}\right)
+\widetilde{R}_{ij} \label{Rij1}%
\end{equation}
where
\begin{equation}
\widetilde{R}_{ij}=\frac{\partial\Gamma_{ik}^{k}}{\partial x^{j}}%
-\frac{\partial\Gamma_{ij}^{k}}{\partial x^{k}}+\Gamma_{im}^{k}\Gamma_{jk}%
^{m}-\Gamma_{ij}^{k}\Gamma_{mk}^{m}\ . \label{Rtil}%
\end{equation}

The general expression valid for the space-space components is
\begin{align}
R_{ij}  &  =\frac{\partial}{\partial r}\left(  \frac{1}{2}\lambda
_{Q}r\widetilde{g}_{rr}\right)  \delta_{ir}\delta_{jr}+\frac{\partial
}{\partial t}\left(  \frac{\overset{\circ}{RR}}{g_{00}}\right)  \widetilde
{g}_{ij}+\left(  \frac{1}{2}\lambda_{Q}\ r\widetilde{g}_{rr}(r)\right)
^{2}\delta_{ir}\delta_{jr}+\frac{\overset{\circ}{R}^{2}}{g_{00}}\widetilde
{g}_{ij}+\nonumber\\
& \nonumber\\
&  +\left(  \frac{1}{2}\frac{\partial\widetilde{g}^{rr}}{\partial r}%
\delta_{ir}\delta_{jr}+\frac{\widetilde{g}^{rr}}{r}\delta_{i\theta}%
\delta_{j\theta}+\frac{\widetilde{g}^{rr}}{r}\delta_{i\varphi}\delta
_{j\varphi}\right)  \widetilde{g}_{ij}\left(  \frac{1}{2}\lambda
_{Q}r\widetilde{g}_{rr}\right)  +\widetilde{R}_{ij}\,\,. \label{Rij3}%
\end{align}

The radial component can be written as
\begin{equation}
R_{rr}=\left(  \frac{R\overset{\circ\circ}{R}+2\overset{\circ}{R}^{2}}{g_{00}%
}\right)  \widetilde{g}_{rr}+\frac{1}{2}\lambda_{Q}\widetilde{g}_{rr}-\frac
{1}{4}\lambda_{Q}r\widetilde{g}_{rr}\left(  \frac{\partial\widetilde{g}^{rr}%
}{\partial r}-\lambda_{Q}\ r\right)  \widetilde{g}_{rr}+\widetilde{R}_{rr}
\label{Rrr}%
\end{equation}
or the alternative form
\begin{equation}
R_{rr}=\left(  \frac{R\overset{\circ\circ}{R}+2\overset{\circ}{R}^{2}}{g_{00}%
}\right)  \widetilde{g}_{rr}+\frac{1}{2}\lambda_{Q}\widetilde{g}_{rr}-\frac
{1}{4}\lambda_{Q}r\widetilde{g}_{rr}\frac{\partial}{\partial r}\ln\left(
\widetilde{g}_{rr}\times g_{00}\right)  +\widetilde{R}_{rr}\,, \label{Rrr1}%
\end{equation}
which contains additional compared with the two angular components. As the
three diagonal space-space components of the Einstein equation must have the
same form, the additional term of the radial equation can be eliminated by
imposing the condition
\begin{equation}
\lambda_{Q}\ r\widetilde{g}_{rr}+\widetilde{g}^{rr}\frac{\partial\widetilde
{g}_{rr}}{\partial r}=0\Leftrightarrow\frac{\partial}{\partial r}\ln\left(
\widetilde{g}_{rr}\times g_{00}\right)  =0\,. \label{condicao}%
\end{equation}

This condition, applied to the time-time component given by equation
(\ref{R00a}) or (\ref{R00b}), leads it to
\begin{equation}
R_{00}=3\frac{\overset{\circ\circ}{R}}{R}+\frac{3}{2R^{2}}\lambda_{Q}%
\ g_{00}\ , \label{R00c}%
\end{equation}
and the space-space components, equations (\ref{Rij3}), assume the general
form
\begin{equation}
R_{ij}=\left(  \frac{R\overset{\circ\circ}{R}+2\overset{\circ}{R}^{2}}{g_{00}%
}\right)  \widetilde{g}_{ij}+\frac{1}{2}\lambda_{Q}\widetilde{g}%
_{ij}+\widetilde{R}_{ij}\ . \label{Rij4}%
\end{equation}

The condition (\ref{condicao}), left equation, can be simplified,
\begin{equation}
\lambda_{Q}\ r\widetilde{g}_{rr}+\widetilde{g}^{rr}\frac{\partial\widetilde
{g}_{rr}}{\partial r}=0\Rightarrow\lambda_{Q}\ r\widetilde{g}_{rr}%
-\widetilde{g}_{rr}\frac{\partial\widetilde{g}^{rr}}{\partial r}=0\ ,
\end{equation}
and rewritten in the simple form
\begin{equation}
\frac{\partial\widetilde{g}^{rr}}{\partial r}-\lambda_{Q}\ r=0\ ,
\end{equation}
its solution given by
\begin{equation}
\widetilde{g}^{rr}=A+\frac{1}{2}\lambda_{Q}r^{2}\ .
\end{equation}
The same condition (\ref{condicao}), right equation,
\begin{equation}
\frac{\partial}{\partial r}\ln\left(  \widetilde{g}_{rr}\times g_{00}\right)
=0
\end{equation}
implies
\begin{equation}
g_{00}=\frac{C}{\widetilde{g}_{rr}}=C\widetilde{g}^{rr}=C\left(  A+\frac{1}%
{2}\lambda_{Q}r^{2}\right)  .
\end{equation}

We are going to impose $A=1$ and $C=-1$ such that, for $\lambda_{Q}=0$, we
will have $\widetilde{g}_{rr}=1$ and $\widetilde{g}_{00}=-1$, as in the
Robertson-Walker metric, equation (\ref{RWalker}), for the case $k=0$. From
such conditions, we get
\begin{equation}
\widetilde{g}_{rr}=\frac{1}{1+\frac{1}{2}\lambda_{Q}r^{2}} \label{grr1}%
\end{equation}
and
\begin{equation}
g_{00}=-\left(  1+\frac{1}{2}\lambda_{Q}r^{2}\right)  , \label{g00b}%
\end{equation}
the metric (\ref{metric}) taking the final form
\begin{equation}
ds^{2}=-d\tau^{2}=-\left(  1+\frac{1}{2}\lambda_{Q}r^{2}\right)  dt^{2}%
+R^{2}(t)\left[  \frac{dr^{2}}{\left(  1+\frac{1}{2}\lambda_{Q}r^{2}\right)
}+r^{2}d\theta^{2}+r^{2}\sin^{2}\theta d\varphi^{2}\right]  \ .
\label{metric1}%
\end{equation}

The scale factor $R(t)$ define the time evolution of the universe, and must be
obtained solving the Einstein equations (\ref{Einstein}). Remember that, to
define completely the space-space components of the Ricci tensor (\ref{Rij4}),
we need to obtain the terms $\widetilde{R}_{ij}$ defined in equation
(\ref{Rtil}), which reduces to
\begin{equation}
\widetilde{R}_{ij}=\frac{1}{r}\frac{\partial\widetilde{g}}{\partial r}%
^{rr}\widetilde{g}_{ij}=\lambda_{Q}\widetilde{g}_{ij}\,, \label{Rtil1}%
\end{equation}
using the components of affine connection given in the appendix. So, equation
(\ref{Rij4}) becomes
\begin{equation}
R_{ij}=\left(  \frac{R\overset{\circ\circ}{R}+2\overset{\circ}{R}^{2}}{g_{00}%
}\right)  \widetilde{g}_{ij}+\frac{3}{2}\lambda_{Q}\widetilde{g}_{ij}\ .
\label{Rij5}%
\end{equation}

The metric (\ref{metric1}) was built imposing just space homogeneity and
isotropy as in the case of the Robertson-Walker metric, but it has a very
important difference between them. Both are for commoving referential, but the
Robertson-Walker was built such that the cosmological time coincides with the
proper time of each observer referential and any observer at rest in any place
will measure the same time. It is possible because all of them are local free
fall referential.. On the other hand, in the charge tempered model, the
commoving referential is not a free fall referential, but instead it is an
accelerated referential, the acceleration due to the Coulomb force, and the
metric (\ref{metric1}) does not permit the synchronization of clocks at
different place. The relation between the time in the commoving referential
and the proper time is given by%

\begin{equation}
-d\tau^{2}=g_{00}dt^{2}\Rightarrow\frac{dt}{d\tau}=\frac{1}{\sqrt{-g_{00}}}\ .
\label{time}%
\end{equation}
Relations between the time derivatives in relation to these two times are
given by
\begin{equation}
\overset{\circ}{R}\ =\frac{\partial R}{\partial x^{0}}=\frac{dR}{d\tau}%
\frac{\partial\tau}{\partial t}=\sqrt{-g_{00}}\frac{dR}{d\tau}=\sqrt{-g_{00}%
}\overset{.}{R} \label{time1}%
\end{equation}
and, as the $g_{00}$ is time independent,
\begin{equation}
\overset{\circ\circ}{R}\ =\sqrt{-g_{00}}\frac{d\overset{\circ}{R}}{d\tau
}=-g_{00}\frac{d^{2}R}{d\tau^{2}}=-g_{00}\overset{..}{R}\,, \label{time2}%
\end{equation}
where
\begin{equation}
\overset{.}{R}\ =\frac{dR}{d\tau}\text{, }\overset{..}{R}\ =\frac{d^{2}%
R}{d\tau^{2}}\text{, etc..} \label{time3}%
\end{equation}
define proper time derivatives. Using such relations, the time-time and
space-space components, (\ref{R00c}) and (\ref{Rij5}) become
\begin{equation}
R_{00}=-3\frac{\overset{..}{R}}{R}g_{00}+\frac{3}{2R^{2}}\lambda_{Q}\ g_{00}
\label{R00d}%
\end{equation}
and
\begin{equation}
R_{ij}=-\left(  R\overset{..}{R}+2\overset{.}{R}^{2}\right)  \widetilde
{g}_{ij}+\frac{3}{2}\lambda_{Q}\widetilde{g}_{ij}\ , \label{Rij6}%
\end{equation}
respectively..

From the time-time component of the Einstein equations (\ref{Einstein}),
\begin{equation}
R_{00}=-8\pi GS_{00}\,, \label{E00}%
\end{equation}
the Ricci tensor component (\ref{R00d}) and the source term (\ref{S00}), we obtain%

\begin{equation}
R\overset{..}{R}\ =\frac{1}{2}\lambda_{Q}\ -\frac{4\pi}{3}GR^{2}(\rho+3p)\ .
\label{acel}%
\end{equation}

From the space-space components of the Einstein equations,
\begin{equation}
R_{ij}=-8\pi GS_{ij}\ ,\label{Eij}%
\end{equation}
which the Ricci tensor and the source components given by (\ref{Rij6}) and
(\ref{Sij}), respectively, results
\begin{equation}
R\overset{..}{R}+2\overset{.}{R}^{2}=\frac{3}{2}\lambda_{Q}+4\pi GR^{2}%
(\rho-p)\,\label{vel}%
\end{equation}
which, combined with the acceleration equation (\ref{acel}) results the
velocity equation
\begin{equation}
H^{2}=\frac{\overset{.}{R}^{2}}{R^{2}}=\frac{1}{2}\frac{\lambda_{Q}}{R^{2}%
}\ +\frac{8\pi}{3}G\rho\ ,\label{Fried}%
\end{equation}
the charge tempered model version of the Friedmann equation. Apart from the
algebraic relations of the equations (\ref{acel}), (\ref{vel}) and
(\ref{Fried}), there is a differential constraint, also, in such a way that
the Friedmann equation (\ref{Fried}) derived once in relation to the time
variable must leads back to equation (\ref{acel}). It implies the differential relation%

\begin{equation}
\frac{\partial}{\partial R}\left(  \rho R^{3}\right)  =-3pR^{2}+\ \frac
{3\lambda_{Q}}{8\pi G}\ \label{diff}%
\end{equation}
or, equivalently,%
\begin{equation}
\frac{\partial(\rho R^{2})}{\partial R}=-\left(  3p+\rho\right)
R+\frac{3\lambda_{Q}}{8\pi GR}\ . \label{diff1}%
\end{equation}

Considering that the charge parameter (\ref{LQ}) is consistent with
$\rho\varpropto R^{-3}$ dependence, that is,
\begin{equation}
\frac{\partial}{\partial R}\left(  \rho R^{3}\right)  =0\Leftrightarrow
\rho=\frac{\rho_{0}}{R^{3}} \label{ro1}%
\end{equation}
one has the pressure equation
\begin{equation}
3pR^{2}=\ \frac{3\lambda_{Q}}{8\pi G}\Leftrightarrow p=\ \frac{\lambda_{Q}%
}{8\pi GR^{2}}\ . \label{p}%
\end{equation}

This last equality, substituted in (\ref{acel}), implies a negative defined
acceleration,%
\begin{equation}
R\overset{..}{R}\ =-\frac{4\pi}{3}GR^{2}\rho<0\ , \label{acel2}%
\end{equation}
and a positive defined deceleration parameter,%
\begin{equation}
q=-\frac{R\overset{..}{R}}{\overset{.}{R}^{2}}=\frac{\frac{4\pi}{3}GR^{2}\rho
}{\frac{1}{2}\lambda_{Q}\ +\frac{8\pi}{3}G\rho R^{2}}>0\ . \label{q}%
\end{equation}
To have a positive acceleration (negative deceleration parameter) it is
necessary positive energy ($k=-1$), not allowed for a charged model, which
demands $k=0$.

\subsection{Solution of Friedmann Equation}

Let us consider the Friedmann equation (\ref{Fried}) as differential equation,%

\begin{equation}
\left(  \frac{dR}{d\tau}\right)  ^{2}=\ \frac{8\pi G\rho_{0}}{3R}%
+\frac{\lambda_{Q}}{2}\ , \label{Fried1}%
\end{equation}
where the charge parameter%
\begin{equation}
\lambda_{Q}=2H_{0}^{2}\left(  1-\Omega_{0}\right)  \label{LQ1}%
\end{equation}
is given in terms of the Hubble constant and the density parameter
$\Omega=\rho/\rho_{c}$ evaluated at present time, $H_{0}$ and $\Omega_{0}%
=\rho_{0}/\rho_{c,0}$, respectively, where the mass density critical value is
defined as usual,%
\[
H^{2}=\frac{8\pi G}{3}\rho_{c}\ .
\]
. Using auxiliary variables%
\begin{equation}
D=\ \frac{16\pi G\rho_{0}}{3\lambda_{Q}}=\ \frac{\Omega_{0}}{\left(
1-\Omega_{0}\right)  } \label{D}%
\end{equation}
and%
\begin{equation}
\text{ }dt^{\prime}=\sqrt{\frac{\lambda_{Q}}{2}}d\tau=H_{0}\sqrt{\left(
1-\Omega_{0}\right)  }d\tau\ , \label{ta}%
\end{equation}
Friedmann equation assumes the simple form%
\begin{equation}
\left(  \frac{dR}{dt^{\prime}}\right)  ^{2}=\ \frac{D}{R}+1\ , \label{Fried2}%
\end{equation}
which parametric solution given by
\begin{equation}
R=\ \frac{\Omega_{0}}{\left(  1-\Omega_{0}\right)  }\sinh^{2}\psi=\frac
{\Omega_{0}}{2\left(  1-\Omega_{0}\right)  }\left(  \cosh2\psi-1\right)
\label{R1}%
\end{equation}
and%
\begin{equation}
t=\frac{1}{H_{0}}\frac{\Omega_{0}}{2\left(  1-\Omega_{0}\right)  ^{3/2}%
}\left(  \sinh2\psi-2\psi\right)  \ . \label{t1}%
\end{equation}

Notice that in this charged environment, while $k=0$, $\Omega_{0}\neq1$, with
a charge dependence on the mass density related by%
\begin{equation}
\rho_{0}=\Omega_{0}\rho_{c,0}\ =\rho_{c,0}-\frac{3}{16\pi G}\lambda_{Q}\ .
\label{ro}%
\end{equation}

For a present matter density (forgetting for a moment the dark energy and dark
matter densities) $\rho_{0}\simeq0.03\rho_{c,0}$ , using $G=6.67259\times
10^{-8}cm^{3}/(g.s^{2})$, $\rho_{c,0}\simeq1.88\times10^{-29}g/cm^{3}$ and
$e=4.8032\times10^{-10}esu=4.8032\times10^{-10}\sqrt{gcm^{3}}/s$,\ one obtain%
\begin{equation}
\lambda_{Q}=\frac{16\pi G}{3}\left(  \rho_{c,0}-\rho_{0}\right)
=2.04\times10^{-35}\times1/s^{2} \label{NL}%
\end{equation}
and%
\begin{equation}
\rho_{Q,0}=2.5\times10^{-24}\frac{e}{cm^{3}}\approx\frac{e}{\left(
1000km\right)  ^{3}}\ . \label{NroQ}%
\end{equation}

An interesting relation can be obtained equations (\ref{LQ}) and (\ref{LQ1}),%
\[
\rho_{Q,0}^{2}=2G\rho_{c,0}^{2}\Omega_{0}\left(  1-\Omega_{0}\right)  \ ,
\]
imposing an upper limit to the charge density,%
\[
\rho_{Q,0}\lesssim\rho_{c,0}\sqrt{G/2}\approx\frac{7\times e}{\left(
1000km\right)  ^{3}}\ ,
\]
where $e$ is the absolute value of the electron charge.

Equations (\ref{R1}) and (\ref{t1}) give us the age of the universe%
\begin{equation}
t_{0}=\frac{1}{H_{0}}\frac{\Omega_{0}}{2\left(  1-\Omega_{0}\right)  ^{3/2}%
}\left(  \sinh2\psi_{0}-2\psi_{0}\right)  \ ,
\end{equation}
where the parameter $\psi_{0}$ is obtained imposing the condition $R(\psi
_{0})=1$,
\begin{equation}
\sinh\psi_{0}=\sqrt{\frac{\left(  1-\Omega_{0}\right)  }{\Omega_{0}}}\ .
\end{equation}

It is to note that as $\Omega_{0}\rightarrow\infty$, the charge distribution
is going to vanish, $\lambda_{Q}\rightarrow0$, the equation (\ref{t1}) going
to%
\begin{equation}
\psi=(1-\Omega_{0})^{1/2}\left(  \frac{3}{2}H_{0}t\ \right)  ^{1/3}%
\end{equation}
in such a way that (\ref{R1}) is going to
\begin{equation}
R(t)=\frac{1}{(1-\Omega_{0})}\psi^{2}=\left(  \frac{3}{2}H_{0}t\ \right)
^{2/3}\ ,
\end{equation}
the usual solution of the Friedmann-Robertson-Walker standard model for the
case $k=0$.

\subsection{Red Shift}

A non relativistic Doppler effect due to the recessional kinetic motion is
given by%
\[
\lambda_{0}=\lambda_{1}\left(  1+\beta_{1}\right)  \ ,
\]
where $\lambda_{0}$ and $\lambda_{1}$ are the observer (at position $r_{0}$)
and the source (at position $r_{1}$) wave length, respectively, and $\beta
_{1}$ is the source recession velocity. Now, there is a gravitational red
shift due to the time dilatation, equation (\ref{time}), which implies%
\[
\lambda_{0}=\frac{\lambda_{1}}{\sqrt{-g_{00}(r_{1})}}\ ,
\]
and, to the first order approximation, it is to be combined as%
\begin{equation}
\frac{\lambda_{0}}{\lambda_{1}}=\frac{1+\beta_{1}}{\sqrt{-g_{00}(r_{1})}%
}\ .\label{Doppler}%
\end{equation}

On the other hand, from (\ref{R(t)}), the radial velocity is given by
\begin{equation}
v(t)=\frac{dr}{dt}=\overset{.}{R}\frac{d\tau}{dt}r_{0}=\sqrt{-g_{00}%
(r)}H(t)\ r(t)\ , \label{vel1}%
\end{equation}
which at the present time is going to be%
\begin{equation}
v(t_{0})=\sqrt{-g_{00}(r)}H_{0}\ r(t_{0})\ , \label{Huble}%
\end{equation}
the Hubble law, now modified by the $g_{00}(r)$ potential term. This potential
term correction, for a charge parameter value given by (\ref{NL}), turn to be
relevant only at a distance near about $Gpc$.

Also, for a light signal travelling from the source at distance $r_{1}$ to the
observer, the first wave front emitted at time $t_{1}$ and the next at the
time $t_{1}+T_{1}$, arriving to the observer at time $t_{0}$ and $t_{0}+T_{0}%
$, respectively, one obtain the relation
\begin{equation}
\frac{T_{1}}{T_{0}}=\frac{\lambda_{1}}{\lambda_{0}}=\frac{R(t_{1})}{R(t_{0}%
)}=R(t_{1})\
\end{equation}
such that the red shift parameter is%
\begin{equation}
z_{1}=\frac{\lambda_{0}-\lambda_{1}}{\lambda_{1}}=\frac{1}{R(t_{1})}%
-1=\frac{1+\beta_{1}}{\sqrt{-g_{00}(r_{1})}}-1\ , \label{z1}%
\end{equation}
with the inverse relation%
\begin{equation}
R(z_{1})=\frac{1}{z_{1}+1}=\frac{\sqrt{-g_{00}(r_{1})}}{1+\beta_{1}%
}\ \label{R(z)}%
\end{equation}
and the velocity as a function of the parameter of red shift,
\begin{subequations}
\begin{equation}
\beta_{1}=\left(  z_{1}+1\right)  \sqrt{-g_{00}(r_{1})}-1\ . \label{beta1}%
\end{equation}
This last equation can be combined with equation (\ref{vel1}), resulting
\end{subequations}
\begin{equation}
z_{1}\ =H_{0}\ r_{1}(t_{0})+\frac{1}{\sqrt{-g_{00}(r_{1})}}-1\simeq
H_{0}\ r_{1}(t_{0})\ . \label{z2}%
\end{equation}

\section{Conclusion}

A charge tempered cosmological model is proposed to describe an universe with
a small charge asymmetry, which excess is distributed uniformly in accordance
to the Cosmological Principle. A very characteristic of this charged model is
a non inertial observer commoving frame of reference, which implies a metric
with a potential term carried by the time-time component of the metric tensor.
Another important feature is that an unique possibility is for a metric with
the curvature parameter $k=0$, in such a way that the amount of any excess of
charge is strongly constrained to the Coulomb force does not surpass the
gravitational one (which can occur just for $k=-1$). As a consequence, there
is not allowed to be responsible of the positive acceleration of the
recessional motion of the universe, as shown by the positive defined
deceleration parameter.

Reduction of the Einstein equations to a modified Friedmann equation is done,
as well as its solution obtained. While $k=0$, it does not mean that the
matter density is equal to the critical one; instead, the small charge
parameter can simulate a $k=-1$ condition, as suggested by the hyperbolic
solution of the Friedmann equation. As a final remark, positive acceleration
can be obtained as usual, introducing the cosmological constant and its
physical counterpart, the dark energy. It remains to be necessary, also, to
correct the problem of the age of the universe, which seems not affected by
the charge asymmetry.

\begin{center}
{\LARGE Appendix}
\end{center}

It contains all of the 40 independent components of the affine connection
\[
\Gamma_{\mu\nu}^{\lambda}=\frac{1}{2}g^{\sigma\lambda}\left\{  \frac{\partial
g_{\nu\sigma}}{\partial x^{\mu}}+\frac{\partial g_{\mu\sigma}}{\partial
x^{\nu}}-\frac{\partial g_{\nu\mu}}{\partial x^{\sigma}}\right\}  :
\]

\begin{quotation}
Components $\Gamma_{\lambda\mu}^{0}$:
\begin{align*}
\Gamma_{00}^{0}  &  =\frac{1}{2}g^{00}\frac{\partial g_{00}}{\partial x^{0}%
}=0\\
\Gamma_{r0}^{0}  &  =\frac{1}{2}g^{00}\frac{\partial g_{00}}{\partial r}%
=\frac{1}{2}\lambda_{Q}\ r\widetilde{g}_{rr}(r)\\
\Gamma_{\theta0}^{0}  &  =\Gamma_{\varphi0}^{0}=0\\
\Gamma_{rr}^{0}  &  =-\frac{1}{2}g^{00}\frac{\partial g_{rr}}{\partial x^{0}%
}=-\frac{\overset{\circ}{RR}}{g_{00}}\widetilde{g}_{rr}\\
\Gamma_{r\theta}^{0}  &  =\Gamma_{r\varphi}^{0}=0\\
\Gamma_{\theta\theta}^{0}  &  =-\frac{1}{2}g^{00}\frac{\partial g_{\theta
\theta}}{\partial x^{0}}=-R\overset{\circ}{R}g^{00}\widetilde{g}_{\theta
\theta}=-\frac{R\overset{\circ}{R}}{g_{00}}\widetilde{g}_{\theta\theta}\\
\Gamma_{\theta\varphi}^{0}  &  =0\\
\Gamma_{\varphi\varphi}^{0}  &  =-\frac{1}{2}g^{00}\frac{\partial
g_{\varphi\varphi}}{\partial x^{0}}=-\frac{R\overset{\circ}{R}}{g_{00}%
}\widetilde{g}_{\varphi\varphi}%
\end{align*}

Components $\Gamma_{\lambda\mu}^{r}$:
\begin{align*}
\Gamma_{00}^{r}  &  =-\frac{1}{2}g^{rr}\frac{\partial g_{00}}{\partial
r}=-\frac{1}{2R^{2}}\widetilde{g}^{rr}\lambda_{Q}\ r\widetilde{g}_{rr}%
g_{00}=-\frac{1}{2R^{2}}\lambda_{Q}\ rg_{00}\\
\Gamma_{r0}^{r}  &  =\ \frac{1}{2}g^{rr}\frac{\partial g_{rr}}{\partial x^{0}%
}=\ \frac{1}{2}\frac{\widetilde{g}^{rr}}{R^{2}}2R\overset{\circ}{R}%
\widetilde{g}_{rr}=\frac{\overset{\circ}{R}}{R}\\
\Gamma_{\theta0}^{r}  &  =\Gamma_{\varphi0}^{r}=0\ \\
\Gamma_{rr}^{r}  &  =\frac{1}{2}g^{rr}\frac{\partial g_{rr}}{\partial
r}=-\frac{1}{2}\frac{\partial\widetilde{g}^{rr}}{\partial r}\widetilde{g}%
_{rr}\\
\Gamma_{\theta r}^{r}  &  =\Gamma_{\varphi r}^{r}=0\\
\Gamma_{\theta\theta}^{r}  &  =-\frac{1}{2}g^{rr}\frac{\partial g_{\theta
\theta}}{\partial r}=-\widetilde{g}^{rr}r=-\frac{\widetilde{g}^{rr}%
\widetilde{g}_{\theta\theta}}{r}\\
\Gamma_{\varphi\theta}^{r}  &  =0\\
\Gamma_{\varphi\varphi}^{r}  &  =-\frac{1}{2}\widetilde{g}^{rr}\frac
{\partial\widetilde{g}_{\varphi\varphi}}{\partial r}=-r\widetilde{g}^{rr}%
\sin^{2}\theta\ =-\frac{\widetilde{g}^{rr}\widetilde{g}_{\varphi\varphi}}{r}%
\end{align*}

Components $\Gamma_{\lambda\mu}^{\theta}$:
\begin{align*}
\Gamma_{00}^{\theta}  &  =-\frac{1}{2}g^{\theta\theta}\frac{\partial g_{00}%
}{\partial\theta}=0\ \\
\Gamma_{\theta0}^{\theta}  &  =\frac{1}{2}g^{\theta\theta}\frac{\partial
g_{\theta\theta}}{\partial x^{0}}=\frac{1}{2}\frac{\widetilde{g}^{\theta
\theta}}{R^{2}}2R\overset{\circ}{R}\widetilde{g}_{\theta\theta}=\frac
{\overset{\circ}{R}}{R}\\
\Gamma_{r0}^{\theta}  &  =\Gamma_{\varphi0}^{\theta}=0\\
\Gamma_{rr}^{\theta}  &  =-\frac{1}{2}g^{\theta\theta}\frac{\partial g_{rr}%
}{\partial\theta}\ =0\\
\Gamma_{\theta r}^{\theta}  &  =\frac{1}{2}g^{\theta\theta}\frac{\partial
g_{\theta\theta}}{\partial r}=\widetilde{g}^{\theta\theta}r\widetilde
{g}_{\theta\theta}=\ \frac{\widetilde{g}^{\theta\theta}\widetilde{g}%
_{\theta\theta}}{r}\ =\frac{\widetilde{g}_{\theta\theta}}{r^{3}}\\
\Gamma_{\varphi r}^{\theta}  &  =\frac{1}{2}g^{\theta\theta}\left\{
\frac{\partial g_{r\theta}}{\partial\varphi}+\frac{\partial g_{\varphi\theta}%
}{\partial r}-\frac{\partial g_{r\varphi}}{\partial\theta}=0\right\} \\
\Gamma_{\theta\theta}^{\theta}  &  =\frac{1}{2}g^{\theta\theta}\frac{\partial
g_{\theta\theta}}{\partial\theta}=0\\
\Gamma_{\varphi\theta}^{\theta}  &  =\frac{1}{2}g^{\theta\theta}\frac{\partial
g_{\theta\theta}}{\partial\varphi}=0\ \\
\Gamma_{\varphi\varphi}^{\theta}  &  =-\frac{1}{2}\widetilde{g}^{\theta\theta
}\frac{\partial\widetilde{g}_{\varphi\varphi}}{\partial\theta}=-\frac
{\cos\theta}{\sin\theta}\widetilde{g}^{\theta\theta}\widetilde{g}%
_{\varphi\varphi}=-\sin\theta\cos\theta
\end{align*}

Components $\Gamma_{jk}^{\varphi}$:
\begin{align*}
\Gamma_{00}^{\varphi}  &  =-\frac{1}{2}g^{\varphi\varphi}\frac{\partial
g_{00}}{\partial\varphi}\ =0\\
\Gamma_{r0}^{\varphi}  &  =\Gamma_{\theta0}^{\varphi}=0\\
\Gamma_{\varphi0}^{\varphi}  &  =\frac{1}{2}g^{\varphi\varphi}\frac{\partial
g_{\varphi\varphi}}{\partial x^{0}}=\frac{\overset{\circ}{R}}{R}\ \\
\Gamma_{rr}^{\varphi}  &  =\Gamma_{\theta r}^{\varphi}=0\\
\Gamma_{\varphi r}^{\varphi}  &  =\frac{1}{2}\widetilde{g}^{\varphi\varphi
}\frac{\partial\widetilde{g}_{\varphi\varphi}}{\partial r}=\widetilde
{g}^{\varphi\varphi}r\sin^{2}\theta\ =\frac{\widetilde{g}^{\varphi\varphi
}\widetilde{g}_{\varphi\varphi}}{r}\\
\Gamma_{r\theta}^{\varphi}  &  =\Gamma_{\theta\theta}^{\varphi}=0\\
\Gamma_{\theta\varphi}^{\varphi}  &  =\frac{1}{2}\widetilde{g}^{\varphi
\varphi}\frac{\partial\widetilde{g}_{\varphi\varphi}}{\partial\theta}%
=\frac{\cos\theta}{\sin\theta}\widetilde{g}^{\varphi\varphi}\widetilde
{g}_{\varphi\varphi}\\
\Gamma_{\varphi\varphi}^{\varphi}  &  =\frac{1}{2}g^{\varphi\varphi}%
\frac{\partial g_{\varphi\varphi}}{\partial\varphi}=0
\end{align*}

\end{quotation}

\end{document}